\documentclass[aip,reprint, showpacs, superscriptaddress]{revtex4-1}
\usepackage{amsmath, graphicx,amssymb,color, bm}
\begin{document}
\title{Measuring local volume fraction, long-wavelength correlations and fractionation in a phase-separating polydisperse fluid}
\date{\today}

\author{J.~J.~Williamson}
\email{johnjosephwilliamson@gmail.com}
\affiliation{Department of Physics, Institute for Soft Matter Synthesis and Metrology, Georgetown University, 37th and O Streets, N.W., Washington, D.C. 20057, USA}
\author{R.~M.~L.~Evans}
\affiliation{School of Mathematics, University of Leeds, Leeds LS2 9JT, UK}


\begin{abstract} 
We dynamically simulate fractionation (partitioning of particle species) during spinodal gas-liquid separation of a size-polydisperse colloid, using polydispersity up to $\sim 40\%$ and a skewed parent size distribution. We introduce a novel coarse-grained Voronoi method to minimise size bias in measuring local volume fraction, along with a variety of spatial correlation functions which detect fractionation without requiring a clear distinction between the phases. These can be applied whether or not a system is phase separated, to determine structural correlations in particle size, and generalise easily to other kinds of polydispersity (charge, shape, etc.). 
We measure fractionation in both mean size and polydispersity between the phases, its \textit{direction} differing between model interaction potentials which are identical in the monodisperse case. These qualitative features are predicted by a perturbative theory requiring only a monodisperse reference as input.
The results show that intricate fractionation takes place almost from the start of phase separation, so can play a role even in nonequilibrium arrested states. The methods for characterisation of inhomogeneous polydisperse systems could in principle be applied to experiment as well as modelling.
\end{abstract}

\maketitle

\section{\label{sec:intro}Introduction}

Polydispersity is pervasive in soft matter. Systems which are polydisperse exhibit a continuous variation among constituent particles in e.g.\ size, charge, or shape -- this is common even in nominally `pure' systems, where one would like to think of the particles all belonging to the same overall species. For example, \textit{length} polydispersity of skin ceramide lipids is thought to be key to their function \cite{Das2013}. Polydispersity of colloids strongly affects phase behaviour \cite{Liddle2011}, and acts to significantly degrade the quality of photonic crystals \cite{Allard2004}. In colloid literature, the polydispersity of some property is $\sigma$, the standard deviation of the distribution in units of its mean. Polydispersity of polymers \cite{Rogosic1996, Warren1999} (usually quantified by a `polydispersity index' which is unity in the monodisperse case) is ubiquitous and can be significantly larger than typical values in colloids. Polydispersity is also present -- intentionally -- in virtually all studies of colloidal glasses \cite{Zargar2013}, but relatively little attention has so far been paid to its role beyond the pragmatic necessity of preventing crystallisation (one recent exception being Ref.~\cite{Zaccarelli2013}). In general the presence of a continuum of different particle species leads to greatly increased complexity relative to the monodisperse case, where each particle is strictly identical.

Polydispersity in particle size is particularly common and perhaps the most widely studied form -- the present work continues in this tradition, focusing on size-polydisperse model colloids. We note from the outset that much of the conceptual apparatus is common to other kinds of polydispersity \cite{Evans2000}. 
 
By now there is a reasonably clear picture of how mild polydispersity affects the phase equilibria of hard spheres and related systems \cite{Evans1998, Evans2000, Sollich2009, Sollich2011, Fasolo2004, Wilding2004, Fantoni2006, Bartlett1999}. The kinetics by which polydisperse systems \textit{approach} equilibrium are, however, just as complex and far less well understood \cite{Zaccarelli2009, Williams2008, Martin2003, Martin2005, Schope2007, Liddle2011, Warren1999, Evans2001}. 

Fractionation (partitioning between phases) of the polydisperse property is a key aspect of polydisperse phase separation and has typically been measured either in equilibrium simulations \cite{Wilding2004} or in experiment after long equilibration time \cite{Evans1998, Erne2005, Fairhurst2004}. There is scant data on how these systems behave whilst evolving towards their fractionated equilibria \cite{Williamson2012, Williamson2012a, Leocmach2013}. This may be especially important where phase separation serves as a route to some nonequilibrium arrested state \cite{Varrato2012,DiMichele2013} -- in such cases, the true compositional equilibrium (in terms of fractionation) may never be reached. In other cases, fractionation is required in order to even access the equilibrium phase \cite{Sollich2011}, so that the dynamics of fractionation directly influences whether the system can equilibrate in any meaningful sense \cite{Liddle2011}. As we will show, actually measuring fractionation in a highly inhomogeneous multiple-phase system that is far from equilibrium (so does not contain macroscopic chunks of each phase) is not a trivial task.

In this work we dynamically simulate gas-liquid separation of a polydisperse fluid and measure in detail the effects of fractionation, building significantly on the findings of Ref.~\cite{Williamson2012}. The use of a skewed size distribution and large polydispersity ($\sigma \sim 40\,\%$) motivates development of a coarse-grained Voronoi method for determining local volume fraction with minimal bias with respect to particle size. We also develop a suite of spatial correlation functions which can test for fractionation effects when the distinction between phases is unclear or absent. 

All these methods can be easily applied to simulations and in principle to experimental data too, given the availability of high-quality microscopy \cite{Zargar2013} and recent developments in particle-sizing \cite{Leocmach2013}. They are not specific to size polydispersity, so can be used in cases such as charge or shape polydispersity. The results shed light on polydispersity's influence in an inhomogeneous, phase-separating fluid. The correlation functions introduced, which do not assume distinct phases, may be useful in determining structural effects of polydispersity even in systems that are \textit{prima facie} homogeneous, such as glasses \cite{Zargar2013}.

\section{\label{sec:model}Model and theory}
\subsection{Model\label{sec:modmod}}

The simulation (described in detail elsewhere \cite{Williamson2012}) contains spherical model colloids with hard cores of mean diameter $\langle d \rangle_\textrm{p} \equiv 1$, and attractive square wells of range $\lambda = 1.15$ and depth $u = 1.82\, k_\textrm{B}T$ (a subcritical dimensionless temperature $T_\textrm{eff} \approx 0.55$ \cite{Liu2005}). The parent volume fraction is set to $\phi_{p} = 0.229$, its critical value in the monodisperse limit. In this limit the fluid is in a region of instability to gas-liquid separation \cite{Pagan2005} -- due to the short attraction range, the resulting coexistence is in turn metastable with respect to crystallisation \cite{Williamson2012a} which is not observed in the simulation timescale. In practice, it is found \cite{Williamson2012} that the polydisperse case to be studied here exhibits spinodal gas-liquid separation just as in the monodisperse limit. This is expected from equilibrium work \cite{Wilding2004};\ even for rather large polydispersity, the gas-liquid binodals (more accurately the cloud and shadow curves into which each binodal splits) are in roughly similar positions in the $\phi$ axis to the monodisperse case as long as volume fraction $\phi$ (as opposed to number density $\rho$) is used as the order parameter.

The diameters of the hard particle cores are taken from a truncated Schulz distribution of parent polydispersity $\sigma_\textrm{p}$. The normalised deviation of particle $i$'s diameter $d_i$ from the mean $\langle d \rangle_\textrm{p} \equiv 1$ is given by $\epsilon_i = (d_i - \langle d \rangle_\textrm{p} )/\langle d \rangle_\textrm{p} $. In contrast to the pseudo-Gaussian particle size distribution used in Ref.~\cite{Williamson2012}, the Schulz has a skew, i.e.\ a nonzero third central moment $\langle \epsilon^3 \rangle_\textrm{p}$. We have employed an upper cutoff at $d=2$ for the distributions, in order to avoid large particles which significantly slow the simulation. The total number of particles is $N= 10000$ and the time unit is the mean time for a free particle of mean diameter to diffuse a unit distance.

We have modelled two possible choices for how the pairwise interaction potential $V(r_{ij})$ depends on the particular sizes of two particles $i$ and $j$ and the distance $r_{ij}$ between their centres \cite{Williamson2012}. In the `scalable' case, the square well range depends multiplicatively on the hard core size:
{\begin{equation}
V_\textrm{{scal}}(r_{ij}) = 
\begin{cases}
\infty & \text{if } r_{ij}\leq d_{ij} \\
-u & \text{if } d_{ij} < r_{ij}\leq \lambda d_{ij}\\
0 & \text{if } r_{ij} > \lambda d_{ij}
\end{cases}~,
\label{eqn:scalablesquarewell}
\end{equation}}
\noindent where $d_{ij} = (d_{i} + d_{j})/2$. In the `nonscalable' case the attraction range depends on the hard core size via an additive constant:
{\begin{equation}
V_\textrm{{non-scal}}(r_{ij}) = 
\begin{cases}
\infty & \text{if } r_{ij}\leq d_{ij} \\
-u & \text{if } d_{ij} < r_{ij}\leq d_{ij} + (\lambda - 1)\\
0 & \text{if } r_{ij} > d_{ij} + (\lambda - 1)\\ 
\end{cases}~.
\label{eqn:nonscalablesquarewell}
\end{equation}}
\noindent In the monodisperse case, $d_{ij} = 1$ for all particle pairs and the definitions become \textit{strictly identical}. In the following we also consider for comparison a hard sphere (HS) fluid;\ in that case, $u = 0$ and again there is no distinction between the potentials.

\subsection{Theory}

As described in Ref.~\cite{Williamson2012}, a perturbative theory for polydispersity can be used to predict fractionation at steady state (metastable) gas-liquid coexistence. The relevant thermodynamic potential for fractionation is $A(\rho) = \rho {d\mu ^{\textrm{ex}} (\epsilon)} / {d\epsilon}$, quantifying the variation in excess chemical potential $\mu ^{\textrm{ex}}$ with scaled particle size deviation $\epsilon$. This tells us `how costly it is' (free-energetically) to increase particle size at a given density $\rho$, in the monodisperse reference case. Then, the fractionation of the $n$th moment between the phases depends on the parent value of the $n+1$th moment like so:
\begin{equation}
[\langle\epsilon ^n \rangle]^\textrm{l}_\textrm{g} = - [A/\rho ]^\textrm{l}_\textrm{g} \langle \epsilon^{n+1}\rangle_\textrm{p} + \mathcal{O}(\epsilon^{n+2})~,
\label{eqn:momentfractionation}
\end{equation}
\noindent where for any quantity $x$ we write $[x]^{\rm l}_{\rm g}\equiv x_{\rm l}-x_{\rm g}$. Subscripts $\textrm{l}$, $\textrm{g}$ indicate evaluation in the liquid and gas phases respectively, and $\textrm{p}$ a quantity evaluated for the parent (overall) distribution.

\subsubsection{Qualitative predictions:\ scalable\label{sec:qualscal}}

The distinction between the scalable and nonscalable model potentials (Eqs.~\ref{eqn:scalablesquarewell} and \ref{eqn:nonscalablesquarewell}) was found, for the current parameters, to switch the sign of $ [A/\rho ]^\textrm{l}_\textrm{g}$ \cite{Williamson2012}. For the scalable potential, $- [A/\rho ]^\textrm{l}_\textrm{g} \sim 5.3$ so the gas is predicted to prefer smaller particles than the liquid ($n=1$ in Eq.~\ref{eqn:momentfractionation}). This was confirmed via simulation in the near-monodisperse regime in Ref.~\cite{Williamson2012}. Eq.~\ref{eqn:momentfractionation} predicts this to hold also for fractionation of variance $\sigma^2 \approx \langle \epsilon^2 \rangle$ induced by nonzero \textit{skew} $\langle \epsilon^3 \rangle_\textrm{p}$ of the parent distribution. Thus the truncated Schulz distribution used here, for which $\langle \epsilon^3 \rangle_\textrm{p} > 0$, should cause the gas phase also to have lower \textit{variance} (hence polydispersity) in the scalable case.

\subsubsection{Qualitative predictions:\ nonscalable\label{sec:qualnonscal}}

For the nonscalable potential, $- [A/\rho ]^\textrm{l}_\textrm{g} \sim -2.2$ so the gas phase should prefer \textit{larger} particles, and be of higher polydispersity. 

The two model potentials used here provide a `switch' to control the equilibrium direction of fractionation -- at least to the extent that the perturbative theory of polydispersity \cite{Evans2000} and the approximate square well free energy \cite{Williamson2012} remain valid. 
In a physical system, interactions will generally be polydisperse in \textit{depth}, not just range. For example, although the depletion attraction in colloid-polymer mixtures somewhat resembles our nonscalable interaction viz.\ the dependence of its \textit{range} on particle size, its attraction strength is also size-dependent, with larger particles experiencing a stronger attraction. In Ref.~\cite{Evans2000} the overall dependence of the depletion potential on particle size led to $- [A/\rho ]^\textrm{l}_\textrm{g} > 0$, so that the liquid prefers larger particles, as for the \textit{scalable} potential in our model. Experimental work on a different colloid-polymer system qualitatively confirms this fractionation direction for the depletion potential \cite{Liddle2014}. Thus, fractionation in any given physical system is sensitive to all details of the polydispersity in the inter-particle potential. The value of the model potentials used here is that they should exhibit opposite fractionation directions to one another, providing a good test-bed for the methods we will develop.

\begin{figure}[floatfix]
\includegraphics[width=8.6cm]{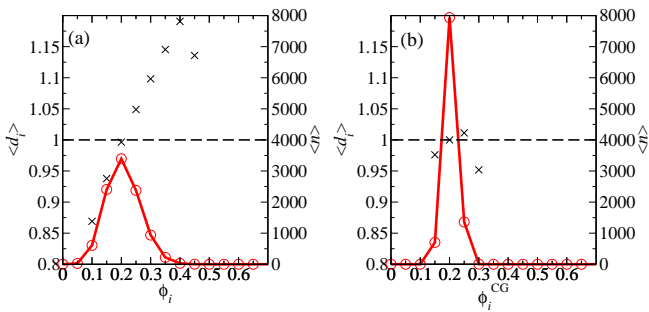}
\caption{\label{naivevsneigh}(a)~Average particle diameter $\langle d_i \rangle$ (black crosses) and number of particles $\langle n\rangle$ (red circles) in bins of the local volume fraction $\phi_i$ in a HS fluid of polydispersity $\sigma_\textrm{p} = 0.0998$. (b)~Using the locally coarse-grained volume fraction $\phi_i ^\textrm{CG}$.
}
\end{figure}

\begin{figure*}[floatfix]
\includegraphics[width=18.6cm]{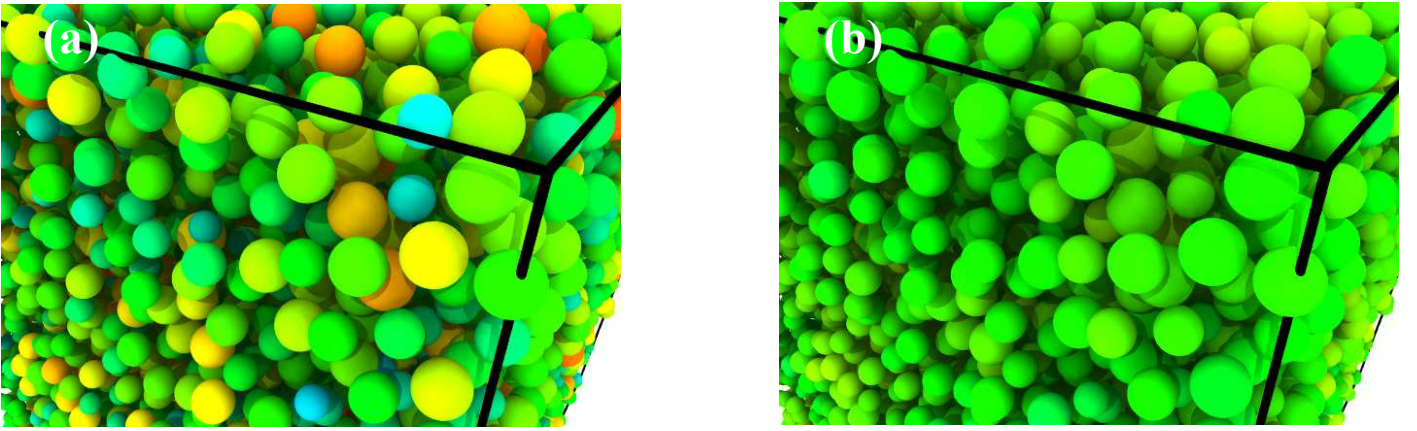}
\caption{\label{HS_combined}(a)~Visualisation of a HS fluid at $\phi_\textrm{p} = 0.229$ and $\sigma_\textrm{p} = 0.0998$ with colours indicating local volume fraction $\phi_i = 0$ (blue) to $\phi_i = 0.4$ (red). It is clear that with this method, high measured volume fraction is correlated with large particle size, although phase separation and thus fractionation is absent. (b)~The same fluid with the locally coarse-grained volume fraction $\phi_i^\textrm{CG}$ plotted instead. Visualisations are made using OVITO \cite{OVITO}.
}
\end{figure*}

\section{\label{sec:results}Results}

The system described in Section~\ref{sec:modmod} is simulated up to $t=8000$, using three independent trajectories for each choice of parameters. We now introduce methods to detect fractionation by locally distinguishing the phases, or examining long-wavelength particle size correlations.

\subsection{Measuring local volume fraction in a polydisperse system\label{sec:measure}}

The obvious way to detect fractionation is by identifying the distinct phases and measuring their properties, requiring a per-particle distinction between the phases. This was done in Ref.~\cite{Williamson2012} by performing a simple neighbour count within a fixed arbitrary range, with a threshold defining which particles are `gas' and which are `liquid'. For larger polydispersities such a method is not satisfactory since, e.g., very large particles may exclude neighbours from their surroundings, leading to an anomalously low neighbour count and tending to cause large particles to be recognised as `gas'. It is not clear how one should adjust the definition to compensate while avoiding simply introducing some other arbitrary bias. 

Therefore one seeks a measure of local density avoiding an arbitrary cutoff. The standard Voronoi cell method is a widely-used solution \cite{Slotterback2008, Voro,Fern2007} but, as we will show, can be significantly improved upon for the present purposes. 

The Voronoi cell is a polyhedron containing all the space that is closer to particle $i$ than any other, and its volume $V^\textrm{Voronoi}_i$ leads to a commonly used definition for $i$'s local volume fraction:
\begin{align}
\phi_i = \frac{V_i}{V^\textrm{Voronoi}_i}~,\label{eqn:naive}
\end{align}
\noindent where $V_i$ is the volume of particle $i$. As well as its widespread use in granular media, the Voronoi cell method has been applied to phase separation of thermal systems, e.g.\ the so-called $2\phi\textrm{MD}$ method in which coexistence conditions are extracted via direct simulation of a two-phase system \cite{Fern2007}. 

\begin{figure}[floatfix]
\includegraphics[width=8.6cm]{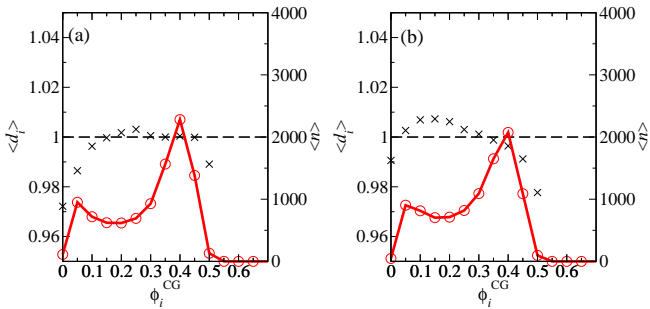}
\caption{\label{svsns}(a)~Average particle diameter (black crosses) and number of particles (red circles) in bins of $\phi_i^\textrm{CG}$ in a phase-separating fluid of polydispersity $\sigma_\textrm{p} = 0.0998$ using the scalable potential. The number peaks corresponding to the gas and liquid are seen, and the gas prefers smaller particles (cf.\ Table~\ref{table:scalfrac}). (b)~Using the nonscalable potential, the fractionation is reversed (cf.\ Table~\ref{table:nonscalfrac}). Error bars are approximately the symbol size. These data are averaged over $t=7200-8000$.
}
\end{figure}

\begin{figure}[floatfix]
\includegraphics[width=8.6cm]{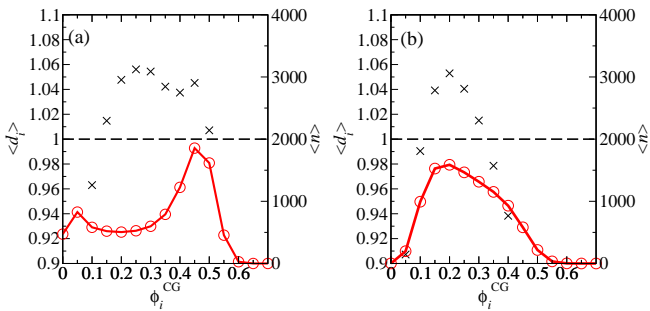}
\caption{\label{40_svsns}As Fig.~\ref{svsns} but for larger polydispersity $\sigma_\textrm{p} = 0.37$.
}
\end{figure}

\begin{figure*}[floatfix]
\includegraphics[width=18.6cm]{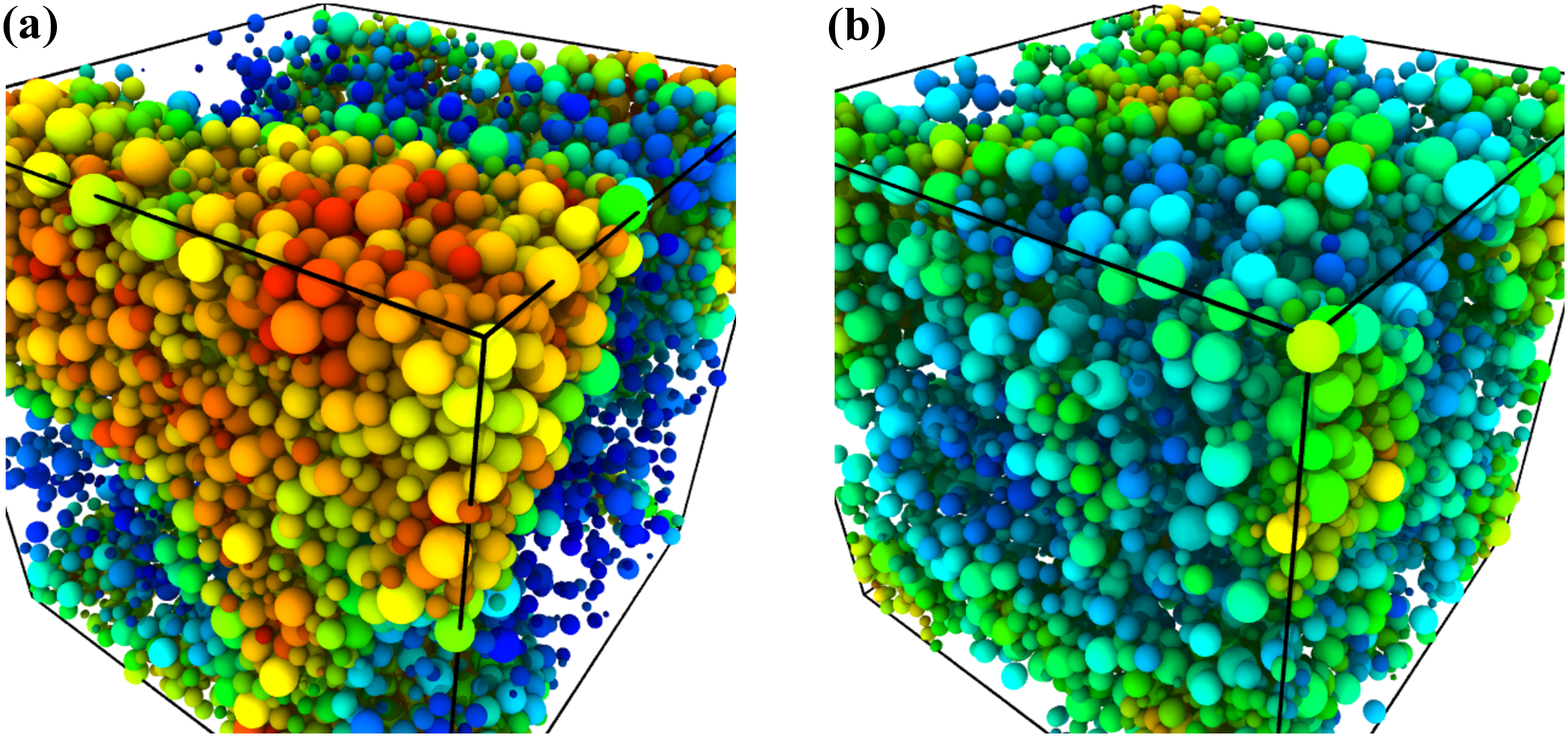}
\caption{\label{40poly_combined}(a)~Visualisation of a gas-liquid separated fluid of polydispersity $\sigma_\textrm{p} = 0.37$ with colours indicating $\phi_i^\textrm{CG} = 0$ (blue) to $\phi_i^\textrm{CG} = 0.6$ (red), using the scalable potential. (b)~Using the nonscalable potential. Particularly for the scalable potential (a), the measured fractionation of smaller particles into the gas can be seen by eye. The snapshots are taken at $t=8000$.
}
\end{figure*}

Given the presence of polydispersity, we use the radical Voronoi tessellation, where boundaries between Voronoi cells are weighted according to relative particle radius, avoiding the problem of Voronoi cell faces `cutting' particles \footnote{It is arguably most natural to first consider the Voronoi `S cell' tessellation, in which the Voronoi cell is defined as the space closer to particle $i$'s \textit{surface} than any other particle's surface, but this causes the cell faces to be curved, considerably increasing computational complexity. The radical tessellation avoids this issue while still closely approximating the structure of the S cells. See Ref.~\cite{Pinheiro2013} for a comprehensive discussion.} (we employ C.~H.~Rycroft's Voro++ library \cite{Voro}). 
However, a difficulty still remains in applying this approach to measure \textit{fractionation}. In Fig.~\ref{naivevsneigh}a we plot the dependence of mean particle diameter on local volume fraction measured with the naive implementation of Eq.~\ref{eqn:naive}, in a hard sphere (HS) fluid of polydispersity $\sigma_\textrm{p} = 0.0998$. Since the HS fluid at this overall volume fraction does not phase separate, a good protocol for our purposes should i)~exhibit a narrow-peaked distribution of particle volume fractions, since all particles are in the same thermodynamic phase;\ ii)~exhibit minimal correlation between particle size and volume fraction, since our purpose later will be to use such a correlation as an indicator of fractionation driven by phase separation, which is absent in the HS fluid. In Fig.~\ref{naivevsneigh}a (using Eq.~\ref{eqn:naive}), we firstly note a wide spread of measured particle volume fractions, reflecting the wide variability in the values of $\phi_i$ generated by Eq.~\ref{eqn:naive}, despite the actual homogeneity of the single-phase system. Secondly, a strong positive correlation is present between particle size and $\phi_i$. This correlation by definition does not represent the phase separation driven fractionation that we intend to study, but a separate `geometric' signal arising from the particular dependence of the denominator versus the numerator of Eq.~\ref{eqn:naive} on particle size in this HS fluid. This presents a significant background bias which interferes with attempting to measure fractionation in a phase-separating system using Eq.~\ref{eqn:naive} -- again it is not clear how to safely subtract this signal out, which would presume existence and knowledge of a consistent relationship between $V_i$ and $V_i^\textrm{Voronoi}$ in the absence of fractionation.

Instead, we revisit the definition of $\phi_i$. Eq.~\ref{eqn:naive} is suitable for getting detailed statistics on the local free volume of particles, but this is not the information one has in mind when speaking about `gas' and `liquid' phases. A particle in a region of liquid randomly experiencing a large upward fluctuation in its own free volume $V^\textrm{Voronoi}_i$ does not thereby become a member of its own one-particle gas phase. A phase in the usual sense is a connected region of space with certain \textit{average} properties, and particles in that region belong to that phase.

Hence, some local coarse-graining is appropriate, but this risks the re-introduction of an arbitrary parameter to set e.g.\ the size of a coarse-graining box. We circumvent this by introducing a coarse-grained volume fraction based on the topology of the Voronoi tessellation:
\begin{align}
\phi^\textrm{CG}_i = \frac{V_i + \sum_j V_j}{V^\textrm{Voronoi}_i + \sum_j V^\textrm{Voronoi}_j}~.\label{eqn:neigh}
\end{align}
\noindent Here, $j$ ranges over neighbours of $i$ where neighbours are those that share a Voronoi cell face. Hence the volume fraction is coarse-grained, but the coarse-graining box is a complex polyhedron defined naturally by the neighbour environment in the vicinity of $i$, avoiding introduction of a new parameter. Fig.~\ref{naivevsneigh}b shows a much more narrow-peaked distribution of volume fractions using this method. The background `geometric' signal correlating particle $i$'s size and local volume fraction is significantly reduced. This is because the local averaging includes the surrounding neighbour region which -- in the absence of fractionation -- will on average contain an unbiased, representative sample of the HS fluid irrespective of $i$'s size.

Comparison of simulation snapshots shows the more uniform local volume fraction when $\phi^\textrm{CG}_i $ is used (Fig.~\ref{HS_combined}b). On the other hand, close inspection of Fig.~\ref{HS_combined}a reveals that those particles of particularly high $\phi_i$ are also the larger ones, reflecting the large bias introduced if Eq.~\ref{eqn:naive} is employed.
As the size of relevant features in a fluid becomes large compared to the particle size (e.g.\ through domain coarsening) and less spatial resolution is required, the averaging could be systematically refined by including neighbours of neighbours (etc.) in the sums in Eq.~\ref{eqn:neigh}. Then, for instance, the proportional contribution of $V_i$ and $V^\textrm{Voronoi}_i$ in Eq.~\ref{eqn:neigh} would be further decreased (thus further reducing the residual size-volume fraction correlation in Fig.~\ref{naivevsneigh}b), at the cost of lower spatial resolution.

As far as we are aware, this is a novel method of defining a locally coarse-grained volume fraction. In the present context it is motivated by the need for a definition that minimises particle size bias present in the absence of fractionation, but the method should be useful in other applications -- including monodisperse systems -- where a local volume fraction measure is required.

\subsection{Fractionation of mean size and polydispersity}

Having chosen $\phi^\textrm{CG}_i $ as a coarse-grained per-particle volume fraction, fractionation can be measured by applying a suitable definition of the gas and liquid phases (dependent on $\phi^\textrm{CG}_i $) and calculating the particle size statistics of the phases. Although the definition of $\phi^\textrm{CG}_i $ is parameter-free, subsequently defining the phases still introduces arbitrary threshold(s). 

In this section we perform fractionation measurements on phase-separating square well fluids, using a simple criterion that assigns $i$ liquid if $\phi^\textrm{CG}_i > 0.229$. Later we will study independent measurements that eliminate even this parameter from the determination of fractionation. The mean diameter and polydispersity of each phase are averaged over $t = 7200-8000$ using truncated Schulz distributions of parent polydispersity $\sigma_\textrm{p} = 0.0998,\, 0.37$. Both the scalable and nonscalable square well potentials are used, which should give opposite fractionation directions (Sections~\ref{sec:qualscal}, \ref{sec:qualnonscal}).

The results are summarised in Tables~\ref{table:scalfrac} and \ref{table:nonscalfrac}. The direction of fractionation is reversed by choosing the nonscalable versus scalable potential. The positive skew $\langle \epsilon^3 \rangle_\textrm{p} > 0 $ of the size distribution leads to fractionation in variance $\sigma^2$ and thus \textit{polydispersity} $\sigma$ between the phases. It is interesting that these important qualitative features hold at large polydispersity, being predicted by a theory that is strictly applicable only in the monodisperse limit. 

We have also expressed the results in the language of Eq.~\ref{eqn:momentfractionation}, which implies $ [\langle \epsilon^2 \rangle ]^\textrm{l}_\textrm{g} / \langle \epsilon^3 \rangle_\textrm{p}  =  [\langle \epsilon \rangle ]^\textrm{l}_\textrm{g} / \langle \epsilon^2 \rangle_\textrm{p}  = - [A/\rho ]^\textrm{l}_\textrm{g} $ at metastable steady state (although we should be careful in placing too much confidence in this prediction far from the monodisperse limit). It appears that the fractionation achieved in the simulation time is weaker than at steady state. Further, we can resolve $[\langle \epsilon^2 \rangle ]^\textrm{l}_\textrm{g} / \langle \epsilon^3 \rangle_\textrm{p}  \neq  [\langle \epsilon \rangle ]^\textrm{l}_\textrm{g} / \langle \epsilon^2 \rangle_\textrm{p} $ in some cases, perhaps implying that the first and second moments are fractionating at proportionally different speeds.
This opens up the interesting possibility of understanding fractionation dynamics in terms of moment relaxation rates \cite{Warren1999}.

Fractionation of polydispersity (as opposed to mean particle size) was detected in experiment \cite{Evans1998} and equilibrium simulation \cite{Wilding2004} but has not previously been measured in the early stages of spinodal decomposition. It may be especially important in cases where the system must form crystals to equilibrate \cite{Liddle2011};\ for instance a liquid phase purified (i.e.\ of reduced $\sigma$) via fractionation may be better able to subsequently nucleate a crystal phase.

Plots of mean diameter and number versus $\phi_i^\textrm{CG}$ (Figs.~\ref{svsns} and \ref{40_svsns}) are in agreement with Tables~\ref{table:scalfrac} and \ref{table:nonscalfrac}. The dominant local volume fractions corresponding to the gas and liquid phases are apparent, and the preference of the gas for smaller particles (scalable potential) or larger particles (nonscalable) can also be seen. Interestingly, at $\sigma_\textrm{p} = 0.37$ nonscalable (Fig.~\ref{40_svsns}b) the distinct peaks in $\langle n \rangle (\phi_i^\textrm{CG})$ are no longer apparent. This is consistent with the less sharply defined phases in Fig.~\ref{40poly_combined}b resulting from slower coarsening, leaving more `interfacial' particles of intermediate $\phi_i^\textrm{CG}$. There may also be an effect of the nonscalable potential on the phase diagram (critical temperature, binodal positions) this far from the monodisperse limit. Fig.~\ref{40poly_combined}b indicates that phase separation is taking place, while Figs.~\ref{quadratic} and \ref{linear} confirm the associated fractionation, but this large polydispersity causes the choice of potential (viz.\ its scaling with polydispersity) to strongly affect structure during phase separation.

\begingroup
\squeezetable
\begin{table}
\begin{ruledtabular}
\begin{tabular}{l | l | l}
$\mathbf{\sigma_\textrm{p}\!=\!0.0998\!\pm\!0.0001}$& Mean diameter & Polydispersity \\
\hline
 Gas & $ \langle d_i \rangle_\textrm{g} = 0.994 \pm 0.001$ & $\sigma_\textrm{g} = 0.0989 \pm 0.0008$ \\ 
 Liquid & $ \langle d_i \rangle_\textrm{l} = 1.0023 \pm 0.0007$ & $\sigma_\textrm{l} = 0.1000 \pm 0.0003$ \\ 
 \hline
$- [A/\rho ]^\textrm{l}_\textrm{g} \sim 5.3$ \cite{Williamson2012}  & $ {\frac{[\langle \epsilon \rangle ]^\textrm{l}_\textrm{g}} { \langle \epsilon^2 \rangle_\textrm{p} }}= 0.84 \pm 0.03$ & ${ \frac{[\langle \epsilon^2 \rangle ]^\textrm{l}_\textrm{g} }{ \langle \epsilon^3 \rangle_\textrm{p} }} =0.8 \pm 0.8 $ \\
 \hline
 & & \\
 \hline
 $\mathbf{\sigma_\textrm{p}\!=\!0.37\!\pm\!0.0008}$&Mean diameter  &Polydispersity  \\
 \hline
Gas & $ \langle d_i \rangle_\textrm{g} = 0.935 \pm 0.004$ & $\sigma_\textrm{g} = 0.341 \pm 0.002$ \\ 
 Liquid & $ \langle d_i \rangle_\textrm{l} = 1.024 \pm 0.003$ & $\sigma_\textrm{l} = 0.377 \pm 0.001$ \\ 
 \hline
 $- [A/\rho ]^\textrm{l}_\textrm{g} \sim 5.3 $ & ${ \frac{[\langle \epsilon \rangle ]^\textrm{l}_\textrm{g} }{\langle \epsilon^2 \rangle_\textrm{p} }}= 0.65 \pm 0.02$ & $ {\frac{[\langle \epsilon^2 \rangle ]^\textrm{l}_\textrm{g} }{ \langle \epsilon^3 \rangle_\textrm{p} }}= 1.04 \pm 0.07$ \\

\end{tabular}
\end{ruledtabular}
\caption{Gas-liquid fractionation results using the scalable square well potential (Eq.~\ref{eqn:scalablesquarewell}). The gas prefers smaller particles and lower polydispersity. Eq.~\ref{eqn:momentfractionation} is tested by comparing ${[\langle \epsilon \rangle ]^\textrm{l}_\textrm{g}  /\langle \epsilon^2 \rangle_\textrm{p} }$ and ${[\langle \epsilon^2 \rangle ]^\textrm{l}_\textrm{g} / \langle \epsilon^3 \rangle_\textrm{p} }$ which, within the perturbative theory \cite{Evans2000}, should both equal $- [A/\rho ]^\textrm{l}_\textrm{g}$ at steady state. It appears that the fractionation measured is weaker than its steady state magnitude. The data are averaged over $t = 7200-8000$.}
\label{table:scalfrac}
\end{table}
\endgroup

\begingroup
\squeezetable
\begin{table}
\begin{ruledtabular}
\begin{tabular}{l | l | l}
$\mathbf{\sigma_\textrm{p}\!=\!0.0998\!\pm\!0.0001}$& Mean diameter & Polydispersity \\
\hline
 Gas & $ \langle d_i \rangle_\textrm{g} = 1.005 \pm 0.001$ & $\sigma_\textrm{g} = 0.1002 \pm 0.0007$ \\ 
 Liquid & $ \langle d_i \rangle_\textrm{l} = 0.9978 \pm 0.0007$ & $\sigma_\textrm{l} = 0.0995 \pm 0.0004$ \\ 
 \hline
$- [A/\rho ]^\textrm{l}_\textrm{g} \sim -2.2$ \cite{Williamson2012}  & $ {\frac{[\langle \epsilon \rangle ]^\textrm{l}_\textrm{g} }{ \langle \epsilon^2 \rangle_\textrm{p} }}= -0.75 \pm 0.04$ & ${ \frac{[\langle \epsilon^2 \rangle ]^\textrm{l}_\textrm{g} }{ \langle \epsilon^3 \rangle_\textrm{p} }}=-0.7 \pm 0.8$ \\

 \hline
 & & \\
 \hline
 $\mathbf{\sigma_\textrm{p}\!=\!0.37\!\pm\!0.0008}$&Mean diameter  & Polydispersity \\
 \hline
Gas & $ \langle d_i \rangle_\textrm{g} = 1.023 \pm 0.004$ & $\sigma_\textrm{g} = 0.373 \pm 0.002$ \\ 
 Liquid & $ \langle d_i \rangle_\textrm{l} = 0.987 \pm 0.003$ & $\sigma_\textrm{l} = 0.368 \pm 0.001$ \\ 
 \hline
$- [A/\rho ]^\textrm{l}_\textrm{g} \sim -2.2$ & $ {\frac{[\langle \epsilon \rangle ]^\textrm{l}_\textrm{g} }{ \langle \epsilon^2 \rangle_\textrm{p} }} = -0.26 \pm 0.02 $ &  ${\frac{ [\langle \epsilon^2 \rangle ]^\textrm{l}_\textrm{g} }{ \langle \epsilon^3 \rangle_\textrm{p} }} = -0.19 \pm 0.04$ \\
\end{tabular}
\end{ruledtabular}
\caption{Gas-liquid fractionation results using the nonscalable square well potential (Eq.~\ref{eqn:nonscalablesquarewell}). The gas now prefers larger particles and higher polydispersity.}
\label{table:nonscalfrac}
\end{table}
\endgroup 

Fig.~\ref{40poly_combined} shows visualisations at $\sigma_\textrm{p} =0.37$ for the two potentials. The coarse-grained volume fraction $\phi_i^\textrm{CG}$ correctly picks out liquid and gas-like regions. Although pictures cannot on their own provide evidence of fractionation, it is satisfying that e.g.\ Fig.~\ref{40poly_combined}a is consistent with the measured fractionation of smaller particles into the gas and larger ones into the liquid.

The results in this section demonstrate that thermodynamically driven fractionation of mean size and also polydispersity (in the presence of a skewed parent distribution) begins during the early stages of gas-liquid separation. The coarse-grained Voronoi method Eq.~\ref{eqn:neigh} provides an improved measure of local volume fraction which substantially reduces particle size bias relative to a naive approach Eq.~\ref{eqn:naive}, enabling definitions of the phases from which fractionation can safely be measured.

\subsection{Real space correlation functions:\ $\xi_2(r)$}

\begin{figure*}[floatfix]
\includegraphics[width=12.6cm]{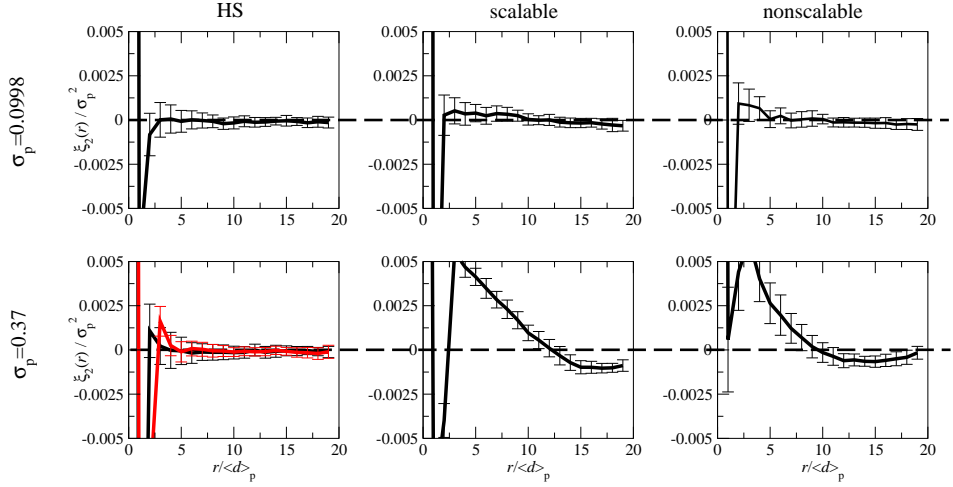}
\caption{\label{quadratic}Size correlation function $\xi_2(r)$ for differing values of the parent polydispersity $\sigma_\textrm{p}$. In the hard sphere (HS) case, there is only a short-range correlation related to local packing efficiency. In the scalable and nonscalable cases, fractionation results in a longer-range correlation on the lengthscale of the gas and liquid domains. This is most easily resolved in the high polydispersity ($\sigma_\textrm{p} = 0.37$) cases. These data are averaged over $t=7200-8000$. The red series is from auxiliary simulations of a HS $\sigma_\textrm{p} = 0.37$ fluid at a higher volume fraction $\phi_\textrm{p} = 0.40$. The only change is an enhancement of the short-range correlation, demonstrating that the long-range signal introduced by phase separation is not simply due to the high volume fraction of the liquid phase.}
\end{figure*}

\begin{figure}[floatfix]
\includegraphics[width=8.0cm]{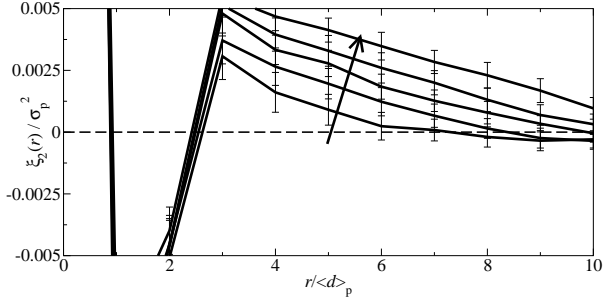}
\caption{\label{time}Evolution of $\xi_2(r)$ at $\sigma_\textrm{p} = 0.37 $ for the scalable potential. The arrow signifies later time intervals:\ $t=0-800,\,800-1600,\,1600-2400,\,2400-3200,\,7200-8000$.
}
\end{figure}

\begin{figure*}[floatfix]
\includegraphics[width=12.6cm]{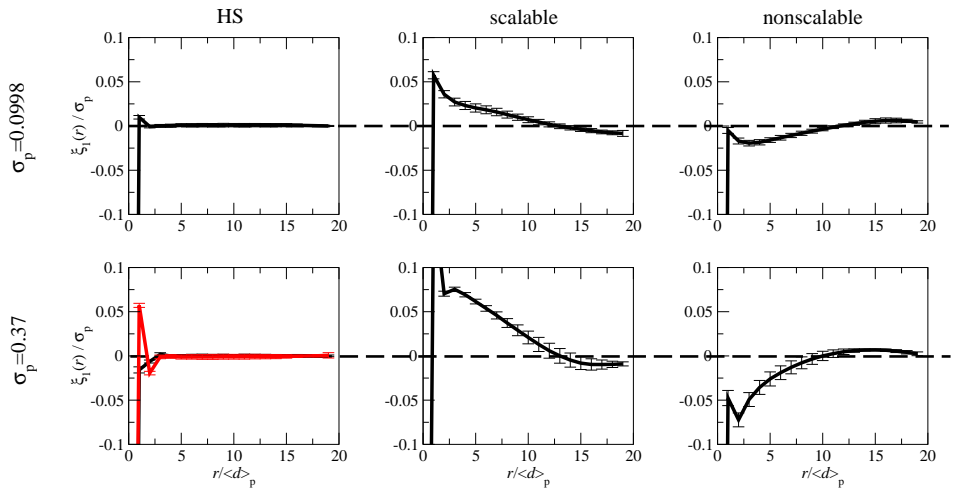}
\caption{\label{linear}As Fig.~\ref{quadratic} but showing the linear size correlation function $\xi_1(r)$. No long-range component is visible in the HS cases. The opposite direction of fractionation in the nonscalable vs.\ scalable case can now be detected. Again, the measurement in a higher $\phi_\textrm{p}$ HS fluid (red series) enhances the short-range correlation but has no effect on the long-range component that signifies fractionation.}
\end{figure*}

Although the coarse-grained volume fraction employed in the previous section was defined in a parameter-free way, subsequently defining the phases inevitably introduces one or more parameters. We are able to corroborate the results with parameter-free methods that detect fractionation not by identifying distinct phases, but by measuring long-wavelength correlations.

If fractionation takes place, we can expect particle size correlations on the lengthscale of the domain size, because particles in the same domain will tend to deviate from $\langle d \rangle_\textrm{p}$ in the same direction -- the signs of their $\epsilon$ will be the same. Such correlations can be tested for using a real space correlation function that resembles a weighted radial distribution function (RDF):
\begin{align}
\xi_2(r) &= \langle \epsilon_i \epsilon_j \rangle_r - \langle \epsilon_i \rangle^2 \\  \notag
&= \left( \frac{\sum_i \sum_{j \neq i} \delta(r_{ij} - r) \epsilon_i \epsilon_j\, }{ \, \sum_k\sum_{l \neq k} \delta(r_{kl} - r)}\right)~
\end{align}
\noindent where $r_{ij}$ is the centre-to-centre distance between particles $i$ and $j$. Note that $ \langle \epsilon_i \rangle \equiv 0 $ since the average is performed over the entire system. $\xi_2(r)$ will be positive whenever particles separated by a distance $r$ are positively correlated in terms of their deviation $\epsilon_i$ from mean diameter. Normalisation by the unweighted $\delta(r_{kl} - r)$ ensures that the structure of the usual RDF is factored out, so $\xi_2(r)$ is only sensitive to correlations in particle size, not in density.

The behaviour of $\xi_2(r)$ is shown in Fig.~\ref{quadratic}. In the HS case, there is only a short-wavelength component reflecting local packing considerations;\ e.g.\ the smallest separations $r$ can only be achieved if both particles involved are small, leading to $\epsilon_i \epsilon_j > 0$ at low $r$. This effect (previously observed in Refs.~\cite{Williamson2012a, Pagonabarraga2000}) is suppressed here due to the large bin size, but we resolve that it is enhanced if the volume fraction of the HS fluid is increased ($\phi_\textrm{p} = 0.4$), as for a standard RDF. Beyond a few particle diameters, particle size is uncorrelated within error, i.e.\ the HS fluids are homogeneous mixtures of size species.

In contrast, measurements in the phase separating square well fluids reveal an additional long-wavelength component. Particles within the same domain are correlated in particle size, signifying fractionation. For $\sigma_\textrm{p} = 0.0998$ this signal is almost too weak to detect within error, but is far stronger for $\sigma_\textrm{p} = 0.37$. 
At $\sigma_\textrm{p} =0.37$, the wavelength of the signal is slightly smaller for the nonscalable potential -- this reflects a smaller characteristic domain size which can also be seen in Figs.~\ref{40poly_combined}b and \ref{unweighted}. Fig.~\ref{time} shows the evolution through time of $\xi_2(r)$, demonstrating growing wavelength of size correlation as the gas and liquid domains coarsen. Finally, we note that the long-range component is absent in the HS case even when volume fraction is increased, demonstrating that it is not simply a result of the high volume fraction of the liquid phase.

\subsection{Real space correlation functions:\ $\xi_1(r)$}

The function $\xi_2(r)$ detects the presence of fractionation, but not its direction. For that purpose we introduce an alternative function which is linear, not quadratic, in the size deviations $ \epsilon $:
\begin{align}
\xi_1(r) &= \langle \epsilon_i + \epsilon_j \rangle_r -2  \langle \epsilon_i \rangle \\  \notag
&= \left( \frac{\sum_i \sum_{j \neq i} \delta(r_{ij} - r) (\epsilon_i +\epsilon_j)\, }{ \, \sum_k \sum_{k \neq l} \delta(r_{kl} - r)}\right)~.
\end{align}
\noindent The `linear' function $\xi_1(r)$ detects fractionation in a different manner to $\xi_2(r)$. Note that the sum over bonds ($\sum_i \sum_{j \neq i}$) will be dominated by the higher density phase, in this case the liquid. If the liquid contains on average larger particles, then for particles in the same domain, $ \epsilon_i +\epsilon_j$ will tend to be positive since particles $i$ and $j$ will, on average, be in the liquid phase. However, if the liquid contains smaller particles, then $ \epsilon_i +\epsilon_j$ will tend to be negative. Hence $\xi_1(r)$ unlike $\xi_2(r)$ is sensitive to the \textit{direction} of fractionation, but relies on a difference in number density between the phases which $\xi_2(r)$ does not.

As shown in Fig.~\ref{linear}, $\xi_1(r)$ like $\xi_2(r)$ decays quickly to zero for HS fluids. The short-wavelength component described in $\xi_2(r)$ is present but manifests differently;\ now the smallest separations $r$ (associated with a pair of small particles) lead to negative $\xi_1(r)$ since $\epsilon_i +\epsilon_j < 0$. In the phase separated case, this dependence on the sign of the $\epsilon$ allows the opposite fractionation directions in the scalable and nonscalable potentials to be resolved. Fractionation of smaller particles into the liquid (nonscalable) results in negative $\xi_1(r)$ for $r$ less than the domain size, as the sum is dominated by liquid particle pairs for which $\epsilon_i +\epsilon_j < 0$. We note also that the relative size of error bars for $\xi_1(r)$ is smaller than for $\xi_2(r)$.

\subsection{Weighted structure factor:\ $S_\xi(q)$}

Since $\xi_2(r)$ and $\xi_1(r)$ contain a short-wavelength component for local packing in addition to the long-wavelength component that signifies fractionation, it is useful to analyse the size correlations in Fourier space using a weighted structure factor:
\begin{align}
S_\xi(\bm{q}) = \frac{1}{N} \langle \sum_{i,j} \epsilon_i \epsilon_j \exp{-i \bm{q} \cdot (\bm{r}_i - \bm{r}_j)}\rangle~.\label{eqn:S}
\end{align}
\noindent In practice we employ the radially-averaged $S_\xi(q)$, and the average over microstates implies, for the phase-separating fluids, an average over the independent trajectories and over suitably short time periods (as for Fig.~\ref{time}). The lowest value of $q$ in each dimension is excluded, to avoid artefacts from the periodic boundary conditions.

In contrast to the real space functions, $S_\xi(q)$ represents particle size correlations in frequency space, so any fractionation signal should be manifest as a size correlation at some low $q$ corresponding to the domain size.

\begin{figure*}[floatfix]
\includegraphics[width=12.6cm]{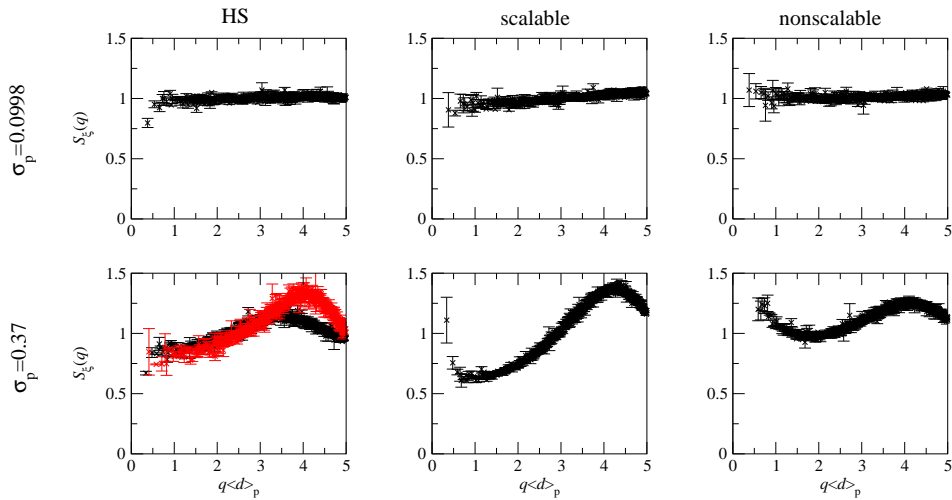}
\caption{\label{weighted}As Fig.~\ref{quadratic} but showing $\epsilon$-weighted structure factor $S_\xi(q)$, which detects the correlations found in $\xi_2(r)$ (Fig.~\ref{quadratic}) in Fourier space. In the HS case, a peak at $q \langle d\rangle_\textrm{p} \sim 4$ corresponds to short-range correlations due to local packing. In the scalable and nonscalable square well cases, fractionation causes an additional peak at low $q$. As for Fig.~\ref{quadratic} these features are clear against noise only in the $\sigma_\textrm{p} = 0.37$ cases, and a higher volume fraction HS system (red series) affects the short-range correlation but does not introduce a low $q$ peak.
}
\end{figure*}

\begin{figure}[floatfix]
\includegraphics[width=7.5cm]{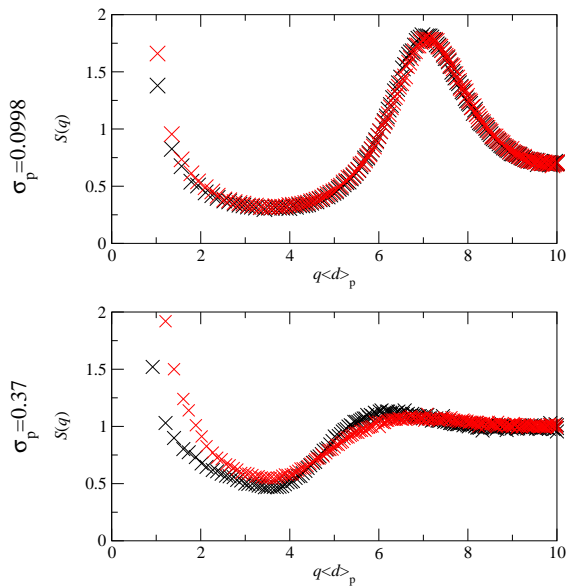}
\caption{\label{unweighted}The standard (unweighted) structure factor $S(q)$ at different values of the parent polydispersity $\sigma_\textrm{p}$ for scalable (black) and nonscalable (red) potentials. Error bars are less than the symbol size.}
\end{figure}

The results are shown in Fig.~\ref{weighted} and are in agreement with the real space measurements in Fig.~\ref{quadratic}. Particularly for $\sigma_\textrm{p} = 0.37$, peaks at low $q$ which are absent in the homogeneous HS case become clear in the phase separating cases. The benefit of $S_\xi(q)$ is that long-wavelength size correlations can be clearly distinguished from short-wavelength packing effects via their separation in $q$.

Comparison of the standard (unweighted) $S(q)$ (Fig.~\ref{unweighted}) reveals interesting differences between the two potentials used. At $\sigma_\textrm{p} = 0.0998$ the two are almost indistinguishable. At $\sigma_\textrm{p} = 0.37$ the low-$q$ peak corresponding to domain formation is at higher $q$ for the nonscalable potential, reflecting the smaller domain size reached on the simulated timescale (Fig.~\ref{40poly_combined}b). Also, the peak corresponding to near-neighbours occurs at higher $q$ for the nonscalable potential. This may be a structural effect of fractionation, the partitioning of smaller particles into the liquid (which dominates the structure factor) causing it to have a smaller inter-particle spacing. Finally, in respect of short-wavelength correlations, the inter-particle peak in $S(q)$ occurs around $q \sim 6$ while its corresponding peak in $S_\xi(q)$ occurred at around $q \sim 4$. This indicates that the size correlations for local packing occur over slightly longer lengthscales than the typical inter-particle distance, and may be related to the phase quadrature (in real space) between size correlation functions and RDFs seen in previous studies of these short-range correlations \cite{Williamson2012a, Pagonabarraga2000}. 

\section{\label{sec:conclusion}Conclusions}

We have investigated ways of robustly measuring structure and correlations in phase separating polydisperse fluids, using them to study fractionation during gas-liquid separation with large polydispersity and a skewed particle size distribution. Fractionation of polydisperse systems can strongly influence the properties of resultant phases \cite{Wilding2004, Sollich2011, Evans1998} and is a necessary process in crystallisation \cite{Sollich2011, Leocmach2013}, the frustration of which is important in glass formation \cite{Zargar2013, Zaccarelli2013}. Like many aspects of polydisperse systems, fractionation is typically studied at equilibrium, and until recently not much was known about the kinetic path towards fractionated equilibrium \cite{Williamson2012, Williamson2012a, Leocmach2013}. Understanding the dynamics of fractionation requires measurement methods which can be applied to inhomogeneous multiple-phase systems, without macroscopic regions of the coexisting phases that can be easily isolated \cite{Evans1998}. 

We have employed a number of such methods to give a comprehensive and consistent picture of fractionation during the early stages of spinodal decomposition, when domains are only a few particle diameters in size. The methods are applicable in principle to experiment as well as simulation, given recent advances in \textit{in situ} particle size measurements \cite{Leocmach2013}. Further, they are easily generalised to polydispersity in properties other than size, in which case $\epsilon_i$ represents an appropriately normalised deviation of the polydisperse property from its mean.

Firstly we introduced a Voronoi method which coarse-grains local volume fraction according to the topology of the Voronoi network. This minimises biasing with respect to particle size, which would otherwise interfere with fractionation measurements. This method can be used wherever a local volume fraction is required (not just in polydisperse systems) such as $2\phi\textrm{MD}$ simulations \cite{Fern2007}, and allows systematic coarsening of the average by including neighbours of neighbours (etc.) without introducing any free parameters. 

We then introduced an alternative approach in which fractionation is detected via associated long-range correlations in particle size (separate from short-range correlations which are sensitive to local packing efficiency \cite{Williamson2012a,Pagonabarraga2000}). In real space, a pair of `weighted RDFs' $\xi_2(r)$ and $\xi_1(r)$ can be used (a related idea occurs in the use of `RDF descriptors' for chemical structure \cite{Hemmer1999,Fernandez2013}). The first shows up fractionation irrespective of how the phases differ, but does not show the direction of fractionation. The second relies on a difference in number density between the phases, but \textit{is} sensitive to the direction of fractionation -- in this case, which phase prefers bigger/smaller particles. It also has smaller characteristic error bars for the system studied here. Finally, it is possible to use a structure factor $S_\xi(q)$ weighted by particle size (or whatever the polydisperse property is) to represent the correlations of $\xi_2(r)$ in Fourier space. This has the advantage of clearly distinguishing the component due to short-range correlations from the long-range one due to fractionation.

With these methods, we measured fractionation of mean diameter and polydispersity, with its direction dependent on fine details of the interaction potential. The important qualitative features are predicted correctly by a theory which only requires as input properties of the monodisperse reference system \cite{Williamson2012, Evans2000}. Given that the system has not reached steady state, it is difficult to judge the quantitative predictions of Eq.~\ref{eqn:momentfractionation} beyond noting that fractionation on the simulated timescale appears not as strong as at predicted steady state;\ one would expect this part-way through phase separation. A direct comparison with fractionation in an equilibrium simulation of the system would shed light on this issue. However, determining polydisperse phase equilibria -- the object of almost all previous work polydispersity -- is itself extremely difficult, requiring highly specialised simulation techniques \cite{Wilding2005} which so far have not been applied to the system studied here. 

Although the fractionation measured here is quite weak, we note that i)~polydispersity of interaction \textit{depth} as well as range will generally exist in a physical system, and can serve to enhance fractionation;\ ii)~our simulations are only in the very earliest stages of phase separation, so we expect that the phases will ripen compositionally on longer timescales \cite{Warren1999}. This latter point could be quantitatively addressed in future work by applying both equilibrium and dynamical simulations to the same system, and comparing the results. 

To avoid significantly slowing the simulation, we truncated the Schulz size distributions at $2 \langle d \rangle _\textrm{p}$. Thus the distributions have lower skew than a true Schulz of the same polydispersity (particularly for the high polydispersity $\sigma_\textrm{p} = 0.37$ case), and we would thus expect stronger fractionation of polydispersity (setting $n=2$ in Eq.~\ref{eqn:momentfractionation}) if the cutoff were removed. Also, we note equilibrium work has shown particle size cutoffs \textit{per se} can have important effects on phase behaviour \cite{Wilding2005}. Fractionation of polydispersity may be very important where crystals nucleate from a metastable liquid \cite{Fortini2008};\ depending on the shape of the parent distribution and the interaction potential, fractionation can reduce or increase the polydispersity of the liquid, thus promoting or suppressing subsequent crystal nucleation. Where crystals are involved, even small changes in polydispersity can have strong effects \cite{Williamson2012a, Martin2003}.

The results show emphatically that complex fractionation is involved right from the beginning of polydisperse phase separation, local composition relaxing alongside local density rather than long after it. It can therefore play a role in the formation of nonequilibrium structures which arrest before coming to equilibrium \cite{Varrato2012,DiMichele2013}, such as when gels form from a polydisperse colloid \cite{Liddle2011}. 

As well as fractionation, the correlation functions introduced could be of general use in characterising structural effects of polydispersity. For instance, polydispersity is ubiquitous in colloidal glasses (in order to prevent crystal growth) \cite{Zargar2013}, but its detailed effects in that context are only just beginning to be understood \cite{Zaccarelli2013}. It would be interesting to study high density amorphous states from the point of view of particle size correlations.

\begin{acknowledgments}

We are grateful for discussions with Dan Blair and Emanuela Del Gado. JJW sincerely acknowledges Chris Rycroft (developer of Voro++ \cite{Voro}) and Alexander Stukowski (OVITO visualisation software \cite{OVITO}) for provision of software and helpful advice. This work was funded by an EPSRC DTG award and by Georgetown University.
\end{acknowledgments}

%

\end{document}